\documentclass[aps,prl,reprint,amsmath]{revtex4-1}
\pdfoutput=1
\usepackage{graphicx}
\usepackage[colorlinks,allcolors=blue]{hyperref}

\begin{document}

\title{Suppression of photon shot noise dephasing in a tunable coupling \\ superconducting qubit}

\author{Gengyan Zhang, Yanbing Liu, James J. Raftery, Andrew A. Houck}
\email{aahouck@princeton.edu}
\affiliation{Department of Electrical Engineering, Princeton University, Princeton, New Jersey 08544, USA.}

\begin{abstract}
We demonstrate the suppression of photon shot noise dephasing in a superconducting qubit by eliminating its dispersive coupling to the  readout cavity. This is achieved in a tunable coupling qubit, where the qubit frequency and coupling rate can be controlled independently. We observe that the coherence time approaches twice the relaxation time and becomes less sensitive to thermal photon noise when the dispersive coupling rate is tuned from several MHz to 22 kHz. This work provides a promising building block in circuit quantum electrodynamics that can hold high coherence and be integrated into larger systems.
\end{abstract}

\maketitle

\section*{Introduction}
Superconducting quantum circuits are a strong candidate for quantum computing \cite{reed2012, lucero2012, corcoles2015} and a convenient platform for quantum optics \cite{houck2007, hofheinz2009, kirchmair2013} and quantum simulation \cite{houck2012, schmidt2013}. Extensive efforts have been made in the last decade to isolate these quantum systems from various decay channels and noise sources in the environment, leading to an increase of several orders of magnitude in energy relaxation time $T_1$ and  phase coherence time $T_2$ (ref.~\onlinecite{devoret2013}). State-of-the-art devices have achieved $T_1$ and $T_2$ in the millisecond regime \cite{reagor2015, pop2014} and pushed gate fidelity close to the threshold for fault-tolerant quantum computing \cite{barends2014}. However, the progress in $T_2$ is slower than that in $T_1$ and $T_2/T_1$ ratios in these devices fall in the range between 0.5 and 1.5 (refs.~\onlinecite{paik2011, rigetti2012, reagor2015}). Deviation from the theoretical limit of $T_2=2T_1$ indicates dephasing mechanisms that need to be understood and circumvented.

In circuit quantum electrodynamics (cQED) \cite{blais2004, wallraff2004}, manipulation and readout of a superconducting qubit are mediated by its coupling to a transmission line cavity. When the coupling is dispersive, photons in the cavity can be utilized to measure the qubit if their phase is shifted by a distinguishable amount depending on the qubit state. On the other hand, changes in cavity photon number will shift the qubit frequency due to the same coupling mechanism. When the amount of the frequency shift is large enough, thermal or quantum fluctuations of cavity photons lead to dephasing of the qubit. This photon shot noise dephasing mechanism has been studied theoretically \cite{gambetta2006} and experimentally \cite{rigetti2012} and has become a dominant factor that limits the coherence time of superconducting qubits. The qualitative discussion above indicates that the dephasing can be suppressed by reducing (1) the photon number fluctuation, characterized by cavity decay rate $\kappa$, (2) the thermal photon population $n_\text{th}$, and (3) the frequency shift caused by each photon, characterized by the dispersive coupling rate $\chi$. Most work in the past has adopted the first two strategies and used high $Q$ ($>10^6$) 3D cavities \cite{sears2012, reagor2015} and careful filtering and thermal anchoring to reduce $\kappa$ and $n_\text{th}$ (refs.~\onlinecite{rigetti2012, jin2015}). Here we focus on the third approach and demonstrate the suppression of photon shot noise dephasing when $\chi$ is tuned to near zero. This is realized in a tunable coupling qubit (TCQ) \cite{gambetta2011, srinivasan2011}, where the qubit frequency and coupling strength can be tuned independently \cite{hoffman2011}. Moreover, we show that measurement of the qubit state can still be performed conveniently when $\chi\ll\kappa$.

In this paper, we start with spectroscopic measurements on the TCQ to demonstrate the independent tunability of its frequency and dispersive coupling rate. We achieve $\chi$ as low as 22 kHz when the dispersive interaction of two different qubit modes is tuned to cancel each other. Next we show that readout of the qubit state near the zero-$\chi$ regime can be realized by utilizing a higher energy level. Finally, we perform time domain measurements of $T_1$ and $T_2$ with injected noise and demonstrate the robustness of $T_2$ against photon shot noise when $\chi$ is near zero.

 \begin{figure*}[htbp]
 \includegraphics{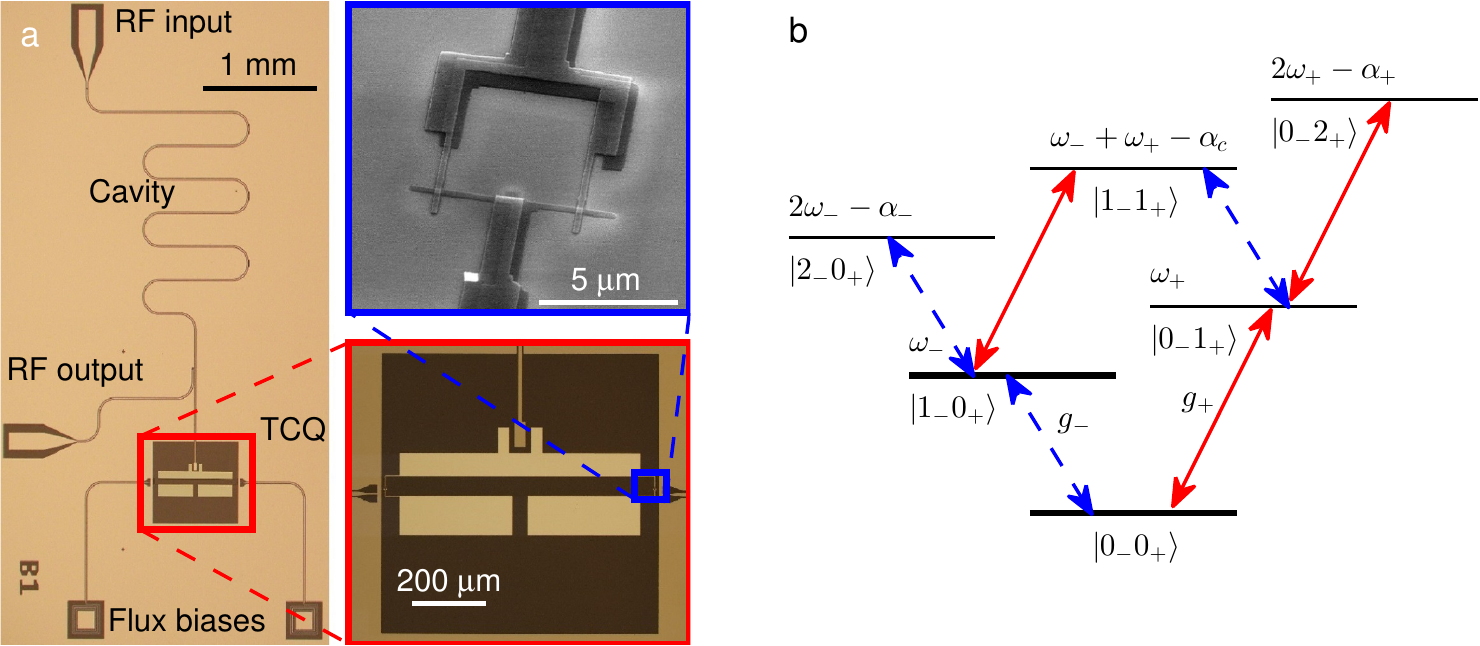}
 \caption{\label{fig:device}TCQ device and its energy levels. (a) Optical micrographs show a TCQ coupled to the end of a readout cavity. RF input and output ports are coupled capacitively to the cavity and spiral inductors are used as low pass filters for the DC bias lines. Electron micrograph shows two Josephson junctions forming a SQUID loop of size $\approx 5 \times 5\ \mu m^2$. (b) V-shaped energy diagram up to two qubit excitations. Red solid and blue dashed lines indicate transitions with dipole coupling rates $g_+$ and $g_-$ respectively.}
 \end{figure*}%
 
\section*{Results}
The TCQ, shown in Fig.~\hyperref[fig:device]{1a}, consists of two transmon qubits \citep{koch2007} strongly coupled to each other via a common third island. The geometry of the islands are designed to minimize the electric participation ratios in the material interfaces and reduce bulk and surface losses of the qubit \cite{dial2015}. The two transmons are formed by capacitively shunted SQUID loops and their frequencies can be tuned by two DC voltages applied to the local magnetic flux bias lines. The strong coupling between the transmon states causes their hybridization and results in a V-shaped energy diagram shown in Fig.~\hyperref[fig:device]{1b}. The one-excitation manifold contains two collective qubit states whose frequencies $\omega_\pm$ and dipole coupling rates $g_\pm$ can be tuned as a function of the two flux biases. Tunable coupling originates from the interference between the dipole moments of the two transmons and provides an extra degree of freedom compared to standard transmon qubits. The two-excitation states acquire different self and cross anharmonicities $\alpha_\pm$ and $\alpha_c$ due to the hybridization, which allows us to drive a desired transition between two states (indicated by arrows in Fig.~\hyperref[fig:device]{1b}) without causing other unwanted transitions. A coplanar waveguide (CPW) cavity with resonance frequency $\omega_r/2\pi = 7.14$ GHz and linewidth $\kappa/2\pi = 250$ kHz is capacitively coupled to the TCQ to drive and read out the qubit states.

We operate the TCQ in the dispersive regime, where the qubit-cavity detunings $|\Delta_\pm|=|\omega_\pm-\omega_r|\gg |g_\pm|$, and the Hamiltonian of the device can be approximated by
\begin{align}
\frac{H}{\hbar}&=\omega_r a^\dagger a + \omega_- b_-^\dagger b_- + \omega_+ b_+^\dagger b_+\notag\\
&\quad+\chi_- a^\dagger a b_-^\dagger b_- + \chi_+ a^\dagger a b_+^\dagger b_+\\
&\quad-\frac{\alpha_-}{2}b_-^\dagger b_-^\dagger b_-b_- - \frac{\alpha_+}{2}b_+^\dagger b_+^\dagger b_+b_+ - \alpha_c b_-^\dagger b_-b_+^\dagger b_+.\notag\label{eqn:hamiltonian}
\end{align}
Here $a,b_\pm$ denote the annihilation operators for the cavity and qubit modes, and $\chi_\pm$ are the dispersive coupling rates between the qubits and the cavity. The measured values for the parameters are $\omega_-/2\pi=7.25$ GHz, $0\lesssim g_-/2\pi<10$ MHz, $\omega_+/2\pi\approx 9.80$ GHz, $g_+/2\pi\approx 90$ MHz, $\alpha_-/2\pi=129$ MHz, $\alpha_+/2\pi=239$ MHz, and $\alpha_c/2\pi=358$ MHz. In this work, we  use the ground ($|0_-0_+\rangle$) and first excited ($|1_-0_+\rangle$) states as the computational basis, so the logical qubit has frequency $\omega_-$ and dispersive coupling rate $\chi_-$. The tunability of $\chi_-$ can be seen from its explicit expression derived in ref.~\onlinecite{gambetta2011} using second order perturbation theory, 
\begin{equation}\begin{aligned}
\chi_- &= \chi_1+\chi_2\\
&=\frac{2g_-^2\alpha_-}{\Delta_-(\alpha_--\Delta_-)}+\frac{g_+^2\alpha_c}{\Delta_+(\alpha_c-\Delta_+)},\label{eqn:chi}
\end{aligned}\end{equation}
where $\chi_{1,2}$ correspond to contributions from the two collective qubit states. The ability to vary $g_\pm$ in addition to $\Delta_\pm$ allows us to tune $\chi_-$ in a flexible way. In particular, when we tune one qubit into the straddling regime \cite{koch2007, inomata2012} and the other far above the cavity, i.e., $g_-\ll\Delta_-<\alpha_-$ and $\Delta_+\gg\max(\alpha_c,g_+)$, $\chi_1$ and $\chi_2$ have opposite signs and $\chi_-$ reaches zero when they cancel each other. In the experiment, we fix $\Delta_-$ and use $g_-$ as the main control knob to tune $\chi_-$: When $g_-$ is large enough, $\chi_1>|\chi_2|$ and $\chi_-$ is positive; As we tune down $g_-$, $\chi_1$ decreases and $\chi_-$ becomes negative when $\chi_1<|\chi_2|$.

\begin{figure}[htbp]
 \includegraphics{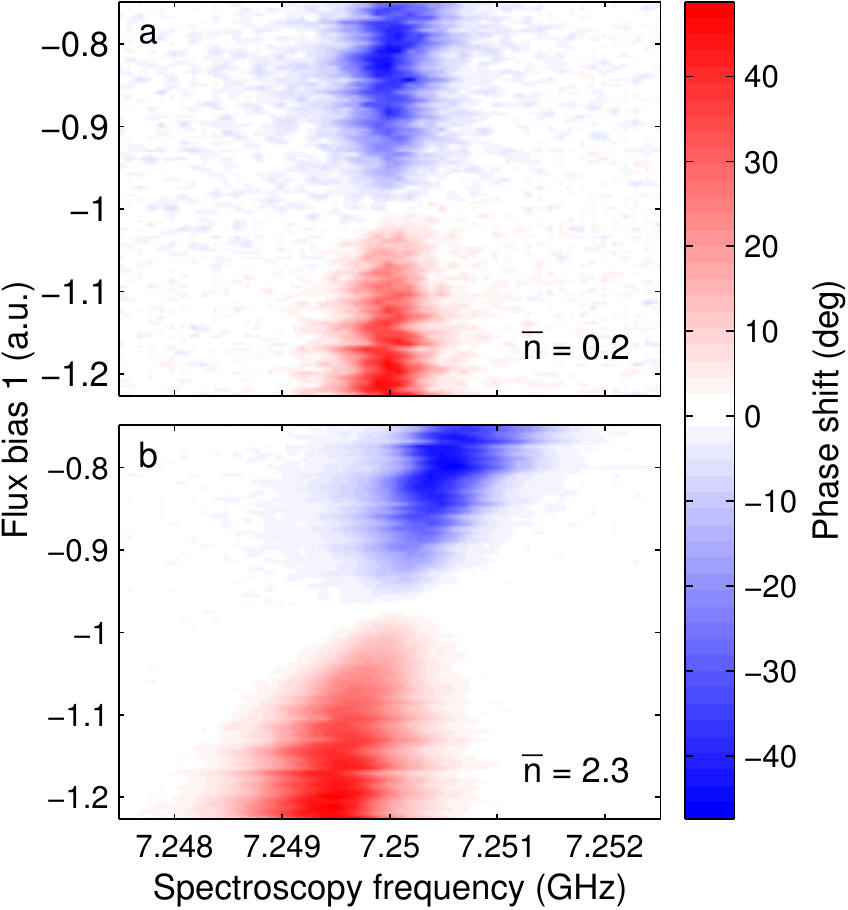}%
 \caption{\label{fig:spec}Tunable dispersive coupling. Phase shift of the cavity transmission is monitored when a second spectroscopy tone sweeps around the qubit frequency. (a) Qubit frequency is fixed at 7.25 GHz when intracavity photon number $\bar{n}=0.2$. The phase shift switches sign when $\chi_-$ crosses zero. (b) When $\bar{n}=2.3$, upward (downward) ac Stark shift of the qubit frequency illustrates positive (negative) $\chi_-$. The spectroscopy signal vanishes near $V_1=-0.95$ V, where $\chi_-$ approaches zero.}
 \end{figure}

Figure \ref{fig:spec} shows the measured data for tuable $\chi_-$. Standard qubit spectroscopy measurement is repeated for different combinations of the two flux biases $(\Phi_1,\Phi_2)$ to map out the constant $\omega_-$ contour in Fig.~\hyperref[fig:spec]{2a}. Here, $\Phi_1$ is varied linearly and $\Phi_2$ (not shown in the figure) is determined by the condition that $\omega_-/2\pi$ remains 7.25 GHz when the intracavity photon number $\bar{n}$ is small. Along the contour,  the phase shift of the cavity transmission changes sign, indicating that $\chi_-$ crosses zero. For larger $\bar{n}$, the qubit frequency is dressed by cavity photons and exhibits an ac Stark shift  \cite{schuster2005} of $\bar{n}\chi_-$. The dressed qubit frequency in Fig.~\hyperref[fig:spec]{2b} shows clearly that $\chi_-$ can be tuned to be both positive and negative, from a few MHz down to below the cavity linewidth $\kappa/2\pi=250$ kHz, which cannot be resolved in the qubit spectroscopy.

\begin{figure}[htbp]
 \includegraphics{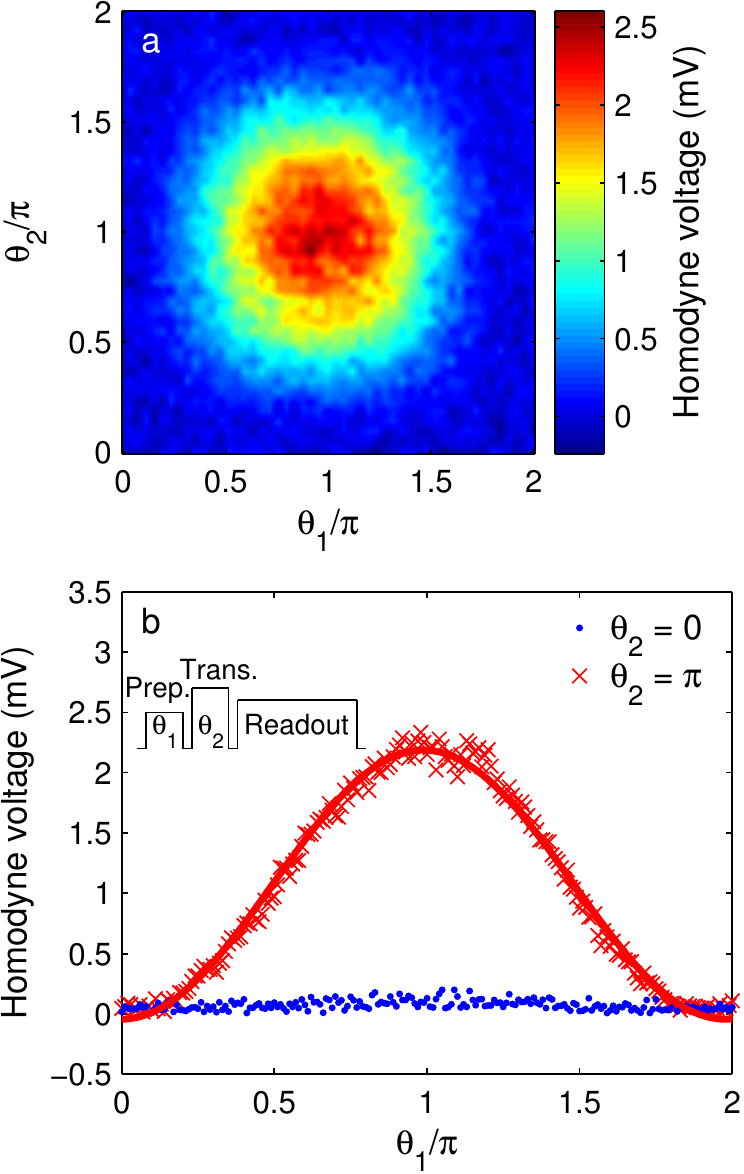}
 \caption{\label{fig:readout}Qubit readout when $\chi_-\approx 0$. Two consecutive Gaussian pulses with $\sigma=16$ ns and frequencies $\omega_-$, $\omega_+-\alpha_c$ are sent to drive qubit transitions, followed by a readout pulse at the cavity frequency $\omega_r$. (a) Homodyne readout signal shows Rabi oscillations for both $|0_-0_+\rangle\leftrightarrow|1_-0_+\rangle$ and $|1_-0_+\rangle\leftrightarrow|1_-1_+\rangle$ transitions, with Rabi angles $\theta_{1,2}$ determined by the amplitudes of the two drive pulses. (b) Horizontal cuts at $\theta_2 = 0$ and $\pi$ in (a). Maximum readout contrast (red crosses) is obtained after transferring $|1_-0_+\rangle$ to $|1_-1_+\rangle$ by a transfer pulse with $\theta_2=\pi$, in contrast to very low readout signal (blue dots) with no transfer pulse. The pulse sequence is shown in the inset.}
 \end{figure}

As $\chi_-$ approaches zero, so does the readout contrast because there is no dispersive shift caused by the $|1_-0_+\rangle$ state, as is illustrated in Fig.~\ref{fig:spec} near $\Phi_1=-0.95$. To achieve efficient readout for small $\chi_-$, we adopt a scheme that utilizes a third state $|1_-1_+\rangle$. In this scheme, we apply a transfer pulse at frequency $\omega_+-\alpha_c$ to the TCQ, inducing a $|1_-0_+\rangle\rightarrow|1_-1_+\rangle$ transition, immediately before the readout pulse at frequency $\omega_r$. The $|1_-1_+\rangle$ state provides a measured dispersive shift of $-1.2$ MHz around the zero $\chi_-$ point and can be used to indirectly read out the logical qubit state. To test this readout method, we prepare the logical qubit to a state $|\psi\rangle$ characterized by Rabi angle $\theta_1$, i.e., $|\psi\rangle = \cos(\theta_1/2)|0_-0_+\rangle+\sin(\theta_1/2)|1_-0_+\rangle$, by sending a Gaussian pulse at frequency $\omega_-$. A transfer pulse with Rabi angle $\theta_2$ is then applied as described above, followed by the readout pulse. The widths of the Gaussian pulses are fixed at $\sigma=$ 16 ns and $\theta_{1,2}$ are controlled by the pulse amplitudes. Figure \hyperref[fig:readout]{3a} shows the measured homodyne signal at $\chi_-\approx 0$ as a function of $\theta_1$ and $\theta_2$. Rabi oscillations for both $|0_-0_+\rangle\leftrightarrow|1_-0_+\rangle$ and $|1_-0_+\rangle\leftrightarrow|1_-1_+\rangle$ transitions are observed, demonstrating coherent transfers between the quantum states. In the absence of the transfer pulse, no visible contrast is observed because of small $\chi_-$ (blue dots in Fig.~\hyperref[fig:readout]{3b}). As the amplitude of the transfer pulse is adjusted to $\theta_2=\pi$, $|\psi\rangle$ is transferred to $|\psi'\rangle=\cos(\theta_1/2)|0_-0_+\rangle+\sin(\theta_1/2)|1_-1_+\rangle$, which gives the maximum readout contrast (red crosses in Fig.~\hyperref[fig:readout]{3b}) and recovers the Rabi angle $\theta_1$. This method allows for the single qubit control and readout scheme to be performed entirely with microwave pulses, and does not involve dynamical tuning of $\chi_-$ via fast flux biasing, which increases experimental complexity and might cause unwanted qubit errors. It also allows low pass filtering to reduce the Purcell decay of the qubit through the flux bias lines \cite{houck2008, reed2010}.

 \begin{figure*}[htbp]
 \includegraphics{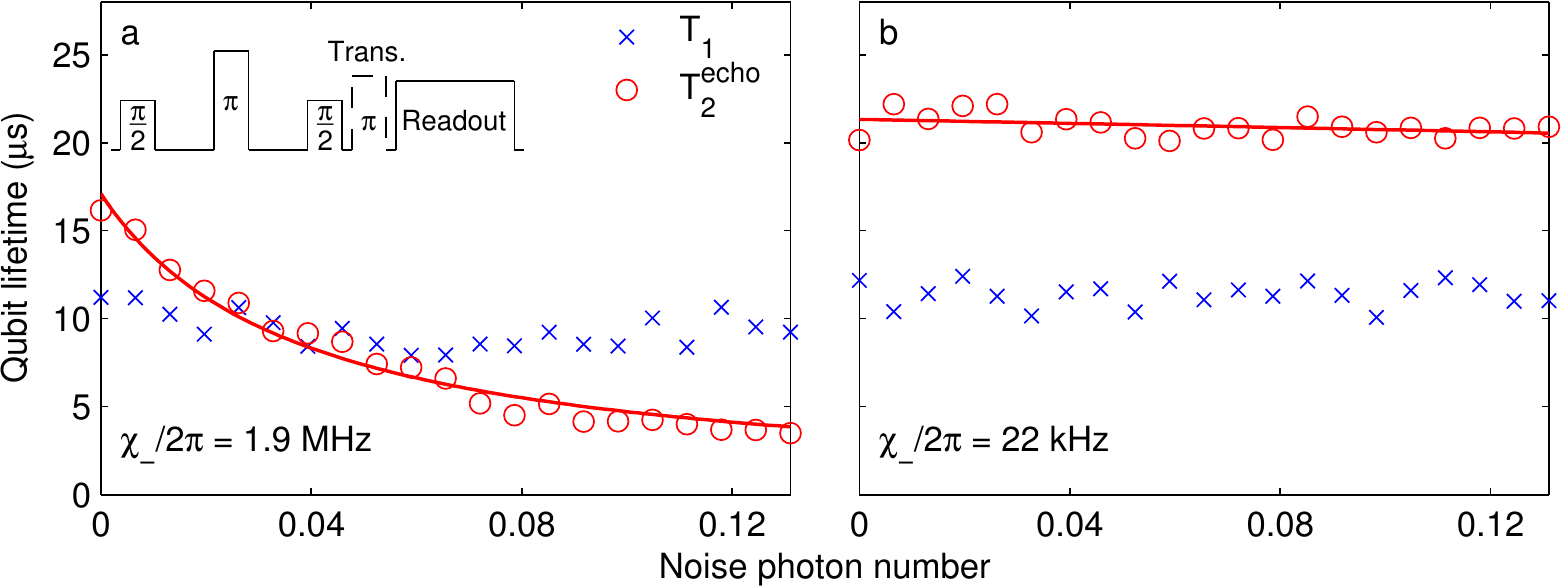}
 \caption{\label{fig:t1t2}Qubit relaxation time $T_1$ (blue crosses), dephasing time $T_2$ with Hahn echo (red dots) and fit to equation (\ref{eqn:dephasingrate}) (red curves) as a function of injected photon noise. (a) When $\chi_-/2\pi=1.9$ MHz, $T_2$ drops from 16 $\mu$s to 3.5 $\mu$s with increasing photon noise. Inset: pulse sequence for $T_2$ measurement with (without) a transfer pulse for small (large) $\chi_-$. (b) When $\chi_-/2\pi=22$ kHz, no drop in $T_2$ is observed up to noise photon number $n_\text{th}$ = 0.13.}
 \end{figure*}%
    
Combining the tunability of $\chi_-$ and the readout method, we perform time domain measurements for the qubit relaxation and coherence time. In addition to the standard measurement setup, we use a noise source to study the influence of thermal photon fluctuations on qubit dephasing. White noise within the bandwidth 7.1375 GHz $\pm$ 5 MHz is injected to the device, covering the cavity but not the qubit, and its power density determines the intracavity noise photon number $n_\text{th}$. In Fig.~\hyperref[fig:t1t2]{4a}, the measured $T_1$ and $T_2$ for $\chi_-/2\pi=1.9$ MHz are plotted as a function of the injected noise power. A Hahn echo pulse is used in measuring $T_2$ to eliminate slow dephasing processes caused by flux noise, etc. While $T_1$ exhibits little dependence on the injected noise and remains $8\sim 11\ \mu$s, $T_2$ drops from 16 $\mu$s to 3.5 $\mu$s as the noise power increases, reflecting a photon shot noise limited $T_2$ when $\chi_-$ is comparable to $\chi$ used in normal transmon devices. Figure \hyperref[fig:t1t2]{4b} shows the result for the same measurements when $\chi_-\approx 0$ and we obtain $T_1= 10\sim 12\ \mu$s and $T_2=20\sim 22\ \mu$s, and no reduction is observed in either $T_1$ or $T_2$ up to $n_\text{th}=0.13$. To quantitatively analyze the result, we use the analytic expression for photon shot noise dephasing rate $\Gamma_\phi$ derived in refs.~\onlinecite{clerk2007, rigetti2012},
\begin{equation}
	\Gamma_\phi = \frac{\kappa}{2}\text{Re}\left[\sqrt{\left(1+\frac{\text{i}\chi_-}{\kappa}\right)^2+\frac{4\text{i}\chi_-n_\text{th}}{\kappa}}-1\right],\label{eqn:dephasingrate}
\end{equation}
and fit the measured $T_2$ data to equation (\ref{eqn:dephasingrate}), shown in the red curves in Fig.~\ref{fig:t1t2}. In Fig.~\hyperref[fig:t1t2]{4a}, a single linear fitting parameter converts the output power of the noise source to the $n_\text{th}$ values in the $x$ axis; In Fig.~\hyperref[fig:t1t2]{4b} we extract $\chi_-/2\pi=22$ kHz from the best fit to equation (\ref{eqn:dephasingrate}). 
 
\section*{Discussion}
To estimate $\Gamma_\phi$, we assume a typical $n_\text{th}=0.02$ (refs.~\onlinecite{rigetti2012,schuster2005}). The small $\chi_-$ leads to $\Gamma_\phi=0.25$ kHz, corresponding to a photon shot noise limited $T_2$ of $4000\ \mu$s. In comparison, to achieve the same level of $\Gamma_\phi$ with the same $n_\text{th}$ in a high $Q$ cavity device would require $\kappa/2\pi=2$ kHz ($Q\sim 3\times 10^6$). The $T_1$ of the device is Purcell limited \cite{houck2008} because of the small detuning between the qubit and cavity, evidenced by a measured $T_1=21\ \mu$s when tuning the qubit to 2.2 GHz below the cavity, and can be improved by increasing cavity $Q$ or engineering the cavity spectrum using filters \cite{reed2010, bronn2015}.

Compared to other methods to suppress photon shot noise dephasing, our approach does not rely on very high $Q$ cavities and $T_2$ is limited by $\chi_-$, which in principle can be tuned to zero. Gate operation and readout can be performed conveniently without dynamical control of the qubit. The planar geometry also makes it easy to integrate the device into larger cQED systems. Potential applications in multi-qubit devices include using the TCQ as coupling bus between two transmon qubits to reduce/enhance their cross coupling strength. The ability to access and measure the quantum state while maintaing the high coherence makes the TCQ a promising building block for the processing and storage of quantum information.

\section*{Methods} 
The device is fabricated on a 500-$\mu$m-thick sapphire substrate. The CPW cavity is defined using photolithography and reactive ion etching of a 200 nm film of niobium sputtered on the sapphire. The TCQ  is patterned using electron beam lithography and the Josephson junctions are made using bridge-free, double-angle evaporation \cite{potts2001}. The chip is mounted and wire bonded to a printed circuit board and cooled down to $\sim 10$ mK in a dilution refrigerator. Input signals generated at room temperature are attenuated and filtered at different stages of the refrigerator. Output signals are amplified at 4 K and room temperature and acquired by a high-speed digitizer.

\section*{Acknowledgements}
This work is supported by IARPA  under  contract W911NF-10-1-0324.

\section*{Competing Interests}
The authors declare no conflict of interest.

\bibliography{zerochi}
 
\end{document}